\documentstyle[twocolumn,prc,aps,epsfig]{revtex}
\input epsf
\newcommand{\be}{\begin{eqnarray}}
\newcommand{\ee}{\end{eqnarray}}
\def\lsim{\mathrel{\rlap{\lower3pt\hbox{\hskip1pt$\sim$}}
     \raise1pt\hbox{$<$}}} 
\def\gsim{\mathrel{\rlap{\lower3pt\hbox{\hskip1pt$\sim$}}
     \raise1pt\hbox{$>$}}} 

\begin{document}

\twocolumn[\hsize\textwidth\columnwidth\hsize\csname @twocolumnfalse\endcsname

\title{Generalized Seniority Description of Cold Fermi Gases}

\author {  G.E. Brown, B.A. Gelman and T.T.S. Kuo}
\address { Department of Physics and Astronomy\\ State University of New York,
     Stony Brook, NY 11794-3800}

\maketitle
\begin{abstract}

We suggest that extension of the Racah seniority description of strongly
interacting fermions in the nuclear shell model is directly generalizable to
describe pairing of atoms in cold Fermi systems. We illustrate this by the
fermionic pairing in the much studied cold two-component gas of $^6\!Li$ atoms.
Our pairing interaction is two orders of magnitude
stronger than that used in the usual
BCS approach. We also explain why the Racah scheme is less applicable
to nuclei, and discuss the similarities of the strongly-coupled matter in cold
fermion systems and the new form of matter found in RHIC close to $T_c$.

\end{abstract}
\vspace{0.1in}
]
\begin{narrowtext}
\newpage
\section{Introduction}

Ultracold atomic gases have become a medium to realize novel phenomena in
condensed matter physics and test many-body theories in new regimes 
\cite{ketterle}.
Recently Regal {\it et al.}
\cite{RGJ} observed a preexisting condensation of fermionic atom 
pairs in the region of magnetic field $B$ above the Feshbach resonance, 
generally referred to as the BCS side. Whereas Regal {\it et al.}
used a trapped gas of fermionic $^{40}\!K$ atoms, Zwierlein {\it et al.}
\cite{ketterle} found
with $^6\!Li$ atoms a much larger condensation of fermionic atoms in the same
BCS region. Such a large condensation of essentially zero momentum molecules
on the BCS side is difficult to understand in the weak coupling BCS theory
following from the Fermi-Thomas approximation usually employed in theoretical
discussions.

By staying in the harmonic oscillator representation (rather then approximating
by plane waves), we utilize the large degeneracy of atoms within a major shell
$N$. This gives a factor of about 60 increase in pairing strength. Pairing is
between time reversed states {\it m} and {\it -m}, giving a simple explanation
for the zero total momentum of each molecule in the condensate.

\section{Generalized seniority}

Racah \cite{racah} 
showed that for a $\delta$-function potential the ground state
solution for the $J^n$ shell was given by coupling all pairs to $J=0$. This is
reasonable, because with fermions, antisymmetry requires that

\begin{equation}
\delta (r_{12}) = - {1\over 3} \vec{\sigma}_1 \, \cdot \,  \vec{\sigma}_2 \,
\delta (r_{12}) \, ,
\label{one}
\end{equation}
so that fermions in one pair will have zero average interaction with those in
another pair. Thus, the energy of a $J^n$ configuration is the sum of pairing
energies (the seniority quantum number being the number of unpaired particles).

Even though the seniority representation gave the exact answer only for an
attractive zero range interaction in the $J^n$ configuration, it could easily
be extended to other degenerate configurations \cite{brown}. 
With a pure pairing interaction 
\begin{equation}
H=- G\, \sum_{m, m^{'}} a_{m}^{\dag} \,  a_{m^{'}}^{\dag} \,  a_{-m} \,
 a_{-m^{'}} \,
\label{Hpair}
\end{equation}
with matrix elements assumed equal for all angular  momenta {\it l} in a major
shell of a principal quantum number $N$, the collective wave function
\begin{equation}
\Psi ={1\over \sqrt{\Omega}}  \left( \begin{array}{c}
 1  \\ 
 1  \\
... \\
 1  \\

\end{array} \right) \, ,
\label{Psi}
\end{equation}
where the states are the paired states of azimuthal quantum numbers
{\it m} and {\it -m} and 
$\Omega$ is the number of pairs $(m, -m)$, is moved downward
in energy with full trace of the secular matrix. This is just the bosonization
of the fermion pairs. We have chosen an example of the $N=15$ shell, with a 
trap energy $15\, \hbar \omega_0$,
of a cold Fermi gas with $^6\!Li$ atoms in mind,
with all levels through $N=15$ being filled. The number of atoms is 1632.

We first make an estimate of the degeneracy. Since in the harmonic trap 
\begin{equation}
\left<r^2\right>_{N {\it l}}=
{\hbar \over m \omega_{0}} (N+{3\over 2}) \, ,
\label{r2}
\end{equation}
then in dimensional analysis, the Slater integral
\begin{equation}
\int R^{2}_{N {\it l}} (r_{1}) R^{2}_{N {\it l}} (r_{2}) \delta (r_{12})
d^3 r_1 d^3 r_2 
\sim \left( {\hbar (N+{3\over 2}) \over m \omega_{0}} \right)^{-3/2} 
\end{equation}
and the total number of states \cite{BM} in shell $N$ is $(N+3/2)^2$ in our
assumption where all Slater integrals, both for ${\it l} = {\it l'}$ and
 ${\it l} \neq {\it l'}$ contribute equally, we find a pairing 
energy\footnote{The coupling constant in eq.~(\ref{Hpair}) is given in terms
of the negative scattering length $a$ by $G=4 \pi \hbar^2 |a|/m$.}
\begin{equation}
\Delta =  \left(N+{3\over 2}\right)^{1/2}\,  
\left( {m \omega_{0} \over \hbar } \right)^{3/2}
\, {4 \pi \hbar^2 |a| \over m}
\end{equation}
so that 
\be
{\Delta \over \hbar \omega_0} = 4 \pi \left(N+{3\over 2}\right)^{1/2}
\, {|a|\over a_{osc}}=51 \,{|a|\over a_{osc}}  \, ,
\label{ourresult}
\ee
where $a_{osc}=(\hbar/m \omega_0)^{1/2}$

We compare this with Heiselberg's eq.~(16) \cite{heiselberg},
\be
{\Delta \over \hbar \omega_0} = {32 \sqrt{2 N_f} \over 15 \pi^2} \, 
{|a|\over a_{osc}} =
1.18 \,{|a|\over a_{osc}} \, ,
\label{Hslb}
\ee
with the last shell $N_f=15$. 

A shell-model calculation, without use of the Thomas-Fermi approximation,
in the
$N_f=12$ shell gave $1.14 |a|/a_{osc}$ to compare with Heiselberg's
$1.06$, showing the Thomas-Fermi approximation to be good.

Because of the large factor between our rough estimate, eq.~(\ref{ourresult}) 
and Heiselberg's, we undertook a numerical diagonilization of the  pairing 
interaction (eq.~(\ref{Hpair})) in the $N=15$ shell without assumption of 
equal matrix elements using harmonic oscillator wave functions. The result was
that the factor 51 in eq.~(\ref{ourresult}) was replaced by 
58.6.\footnote{This is an underestimate because backward-going ladders 
(ground-state correlations) which we have not included give additional 
attraction.} We note that this latter number is $4 \pi$ times $58\%$ of the
trace of the 
matrix we diagonalized. Using the Brown-Bolsterli model \cite{BB}, in which
off-diagonal matrix elements are approximated by
\be
M_{ij}=\sqrt{M_{ii}\, M_{jj}} \, ,
\ee
the factor would be $4\pi$ times the trace of the matrix, so that, within
factor of about 2, the Brown-Bolsterli factorisable model, which involves 
calculation of only the diagonal matrix elements, can be used. This should
facilitate calculations involving many shells. 

From the similarity of the mathematics with the Brown-Bolsterli model, we see
that it is the degeneracy of the configurations within the shell of a given
principle quantum number which gives us our factor of about 60
as compared with weak coupling BCS. In fact, as can be seen,
the usual BCS treatment gives a rather weak pairing, with a gap
\be
\Delta \approx E_f \, \rm{exp\left(-{\pi \over 2 |a| k_f}\right)} 
\ee
aside from a factor of about unity. In general, $k_f$ is taken as 
$\sqrt{2 N_f}/a_{osc}$, where $N_f = 15$ in our case. This pairing is a surface
effect, holes being made over a region of momenta about the Fermi surface so
that particles of equal magnitude, but opposite momenta, can scatter with
them.

In our case of a volume effect, the gap begins linearly with $a$, increasing
rapidly with increasing degeneracy. With the large
binding energy of eq.~(\ref{ourresult}) we can easily increase $\Delta$ 
so that it cancels the trap energy for the $N_f=15$ shell. In fact, we can
make the pair of fermions into a boson,
following a scenario somewhat similar to that suggested by Falco and Stoof
\cite{stoof}. The molecular binding energy of the boson is
\be
{\hbar^2 \over m\, a^2}= \hbar \omega_{0} \, \left({a_{osc}\over a}\right)^2
\,.
\ee 

We can provide the total binding energy from the mean field energy on
the right hand side of eq.~(\ref{ourresult}):
\be
30 + \left({a_{osc}\over a}\right)^2 =4\pi \,
\left(N_f +{3\over 2}\right)^{1/2} \, {|a|\over a_{osc}} \, ,
\ee
which gives 
\be
{a_{osc}\over |a|}=1.75 \,.
\ee
Using the $k_f=\sqrt{2 N_f}/a_{osc}$ \cite{heiselberg} this gives
\be
{1\over k_f \, |a|} = 0.32 \, .
\ee
For $8 \times 1632$ atoms $(k_f \, |a|)^{-1}$ would be about half of this.

Our factor 60 increase is achieved chiefly by correlation in angle. By
pairing the {\it m} and {\it -m} projections of angular momentum and summing
over {\it m} one gets using the addition theorem for angular momenta
$$\Psi_{\rm{L}=0} \left(\theta, \phi \right)=
\sqrt{2 {\it l} +1}\, P_{\it l} \left(\rm{cos}\, \theta_{12} \right) $$
for the angular dependence of the paired state \cite{brown}. The factors
$\sqrt{2 {\it l} +1}$ were used in the Brown-Bolsterli model \cite{BB}
which was formulated in the representation of good angular momentum;
in the {\it m} representation the matrix in eq.~(\ref{Hpair}) with 
all matrix elements equal is equivalent.

Of course, the $\delta$-function interaction sets
$ P_{\it l} \left(\rm{cos}\, \theta_{12} \right)=1$ so that the factor
$(2 {\it l} +1)$ from squaring $\Psi_{\rm{L}=0}$ resulted in Racah's
seniority scheme. We gain additionally by including the off-diagonal
Slater integrals.

In fact, our interaction is increased by the two-body correlations, so that
$k_f$, which is determined by the total number of particles divided by the
volume of the trap; {\it i.e.}, by the average density, should still be given
by $\sqrt{2 N_f}/a_{osc}$. We can, therefore, rewrite our 
eq.~(\ref{ourresult}) for $\Delta$ as
\be
{\Delta \over \hbar \omega_0} = 11 \, k_f \, |a| \, ,
\ee
neglecting the $3/2$ as compared with $N$, so that the $N$ dependence is
taken up in $k_f$. In this way we do not have to know $a_{osc}$. The 
$\sqrt{2 N_f}$ times 11 gives the factor of about 60 in our enhancement in
the $N=15$ shell.

It should be noted that for the spherical trap we have assumed that 34 out of
this 60 comes from the 31 ${\it l}=15$ pairs and that our seniority 0
solution is exact for a $\delta$-function interaction \cite{racah}. In the
experiments the traps are not spherical, typically being cigar shaped with
quite large ratios of frequencies. In nuclear physics, although with lesser
ratios, one uses the Nilsson model, with
$\bar{\omega} = \left(\omega_x \, \omega_y \, \omega_z \right)^{1/3}$. 
Since, $\left<r^2\right>^{1/2}_{N {\it l}}$, in eq.~(\ref{r2}), goes as
$\omega^{-1/2}_{0}$ this means that the dependence of 
$\left<r^2\right>^{1/2}_{N {\it l}}$ on any particular $\omega_i$, say
$\omega_z$, goes as $\omega^{-1/6}_{z}$. Now in a cigar-shaped
trap there will still be the correlation in $\theta_{12}$ enhancing
the densities of time-reversed orbits {\it m} and {\it -m}, but, especially
at the equator, the distance $r$ from the center of the trap to the edge of
the cigar will be smaller than that of a spherical trap with the same number
of {\it m} values. Thus, the distance $r\, \theta_{12}$ will be shorter
than in the spherical trap, so the enhancement in probability will be
greater. However, this should not be a large effect because of the weak
dependence of $\left<r^2\right>^{1/2}_{N {\it l}}$ on any
particular $\omega_i$. 

For small values of $|a|$, the particles in the levels below $N=15$ will
interact with those in $N=15$ so as to push the level up (although in
the middle levels they will push the levels above up and the those below
down). However, the attraction goes as $\sqrt{N}$, as can be seen from 
eq.~(\ref{ourresult}), whereas the trap energy goes linearly with $N$, which
decreases faster than the attraction with decreasing $N$, so that the lower
levels will have already gone into molecules, the atoms from shell $N_f$
being the last to make the transition.

We have not done dynamics, but we believe that for very small $|a|$ there will
be a collisionless regime, in which the RPA vibrations of Bruun and Mottelson
\cite{bruunmott} occur.
The binding energy of the system, the sum of bubbles which make up
the RPA vibrations or zero-point energy of the vibrations \cite{browngiant}
can be
incorporated, as noted by Bruun and Mottelson, as a mean field which slightly
increases the oscillator energy spacings. But then as $a$ becomes larger
in magnitude the inner shells start losing their support  as the attractive
field from the interaction reaches the trap energy in magnitude and the shells
start collapsing in what becomes a hydrodynamical (strongly interacting)
regime. The end result, already early on for negative values of 
$(k_f\,a)^{-1}$ is
that the strongly bound molecules are formed. The higher shells of filled
atoms may remain, but their $\hbar \omega_0$ will be lowered by the attractive
mean field.

\section{Seniority in Nuclei}

We explain why nuclei, for which Racah introduced seniority,
do not look like the cold Fermi gases we have been
discussing. Although seniority was very useful in discussion of the energies
of nuclei, it has turned out that the short range attraction in nuclei
is totally obliterated by the short-range repulsion, so that the effective
interaction in s-wave is only the long-range part of the potential, from
about $1\, fm$ outwards. This is most clearly understood in the
Moszkowski-Scott separation method in the calculations of Holt and Brown
\cite{brownholt}. 
In the calculation of nuclear interactions, as done by Kuo and
Brown \cite{brownkuo}
which used this separation method, the inner part of the s-wave
two-body interaction up to separation distance $d\sim 1\, fm$ is simply
removed, leaving the long-range part of the two-body interaction $V_{\it l}$
to be used in shell-model calculations. Holt and Brown showed that this 
 $V_{\it l}$ is the configuration space form of  $V_{low-k}$, which is the more
effective effective field theory (MEEFT) interaction and contains all
experimental
information in the nucleon-nucleon scattering phase shifts \cite{vlowk}. This
long range $V_{\it l}$ is clearly more suitable for shell-model calculations
than for pairing.  The latter in nuclei is of the weak coupling BCS type
and takes place only in the last shells, near the Fermi energy.
 
\section{Relation to RHIC Physics}

Edward Shuryak \cite{shuryak}
has often emphasized that the elliptic flow seen by
O'Hara {\it et al.} \cite{ohara}
of the cold Fermi liquid, when released from the trap,
is similar to that formed at RHIC, although there are many orders of
magnitude difference in them ({\it e.g} microseconds versus fermi/c). Our
description of the pairing involves the same wave function $\Psi$ 
(eq.~(\ref{Psi}) as that used by Brown, Lee, Rho and Shuryak 
\cite{BLRS}, (Appendix A. in \cite{BLR}) , although at
RHIC the bosonization of quarks and antiquarks into chirally restored mesons
above $T_c$ is involved, rather than fermion pairs. There the quark and 
antiquark masses from lattice gauge calculations are found to be 
$\sim 1\, GeV$, and the $\pi$ and $\sigma$ meson masses must be brought to
zero at $T=T_c$, in the chiral limit, which requires extremely strong
interactions. The attractive interactions are built up enormously through the
degeneracies.

Thus, although the Brown-Bolsterli model for giant resonances in nuclei
suggested the use of the degeneracies both here and in RHIC,
they are immensely strong, because the numbers are
larger and the degeneracy is not partially broken by spin-orbit coupling, as
it is in nuclear physics. Inclusion of the degeneracy brings about the 
strong coupling.

\section{Discussion}

Our result is consistent with the results of Carlson {\it et al.} 
\cite{carlson}
who found $\Delta$ to be about half of the Fermi energy in quantum Monte Carlo
studies of superfluid Fermi gases. In \cite{BLR}
Brown {\it et al.} discuss fluid formed in RHIC in terms of Nambu Jona-Lasinio
theory. They find that the interactions above $T_c$ are much stronger,
because of the large degeneracy, than below $T_c$, where the unperturbed
spectrum of quarks and quark-holes is continuous as in the Fermi 
gas.\footnote{In a longer paper we shall show schematizing
further the schematic model of a Cooper pair \cite{BCS} that it is possible
to go smoothly from the weak coupling BCS to our strong coupling one,
as the degeneracy is introduced.}

We believe the experiments of Regal {\it et al.} \cite{RGJ} and
Zwierlein {\it et al.} \cite{ketterle} found
pairs of atoms with properties such as our scenario would produce.
First of all, our molecules are formed on the BCS side so one would
expect large condensate fractions there when the temperature was lowered.
They find a large fraction of molecules to be in a zero-momentum state
after fast ramping in magnetic field over to the BEC side,
``this means that the nearest neighbors had opposite momenta'' 
\cite{ketterle}. In our seniority scenario, atoms of time-reversed states 
${\it m}$ and ${\it -m}$ are paired, states of equal in magnitude and
opposite in
direction angular momenta. In the long-range Cooper pairs discussed in
other works one would expect
the transfer into a tightly bound molecular state would randomly pick out one
of the nearest neighbors, resulting in a thermal molecule.  
Zwierlein {\it et al.} \cite{ketterle} 
regard their high condensate fraction as evidence  
of condensed atomic pairs in the BCS sector which are smaller in size
than the inter atomic distance and, therefore, molecular. The size of
our pairing modes $\Psi$, eq.~(\ref{Psi}), decreases roughly inversely with
increasing $\Omega$, the number coupled together, forming a highly
compact molecular state.
 
The condensate drains particles from the Fermi sea on the BCS side
as we have outlined, but only after they are tightly paired. They go into
zero-momentum molecules so that they are easily Bose condensed. As noted in 
\cite{ketterle} both there and in \cite{jochim} and in \cite{bart} a reduction
in the size of the atomic cloud was observed as the Feshbach resonance was
approached from the BCS side.

We believe that our generalized seniority scenario explains the important
features of the qualitative change in pairing phenomena demanded by the 
Regal {\it et al.} \cite{RGJ} and 
Zwierlein {\it et al.} \cite{ketterle} results. 

Our conclusion is that the generalized seniority description gives
a new strong-coupling pairing which describes important features found
in experiments.

We end by noting that Leggett's ansatz \cite{Leggett} is
\be
\Psi \left(r_{1} \, \sigma_{1} \, ... \, r_{N} \, \sigma_{N} \right) =
A\, \phi \left(r_{1} \, \sigma_{1} \, ,r_{2} \, \sigma_{2} \right) \,
\nonumber \\
\phi \left(r_{3} \, \sigma_{3} \, ,r_{4} \, \sigma_{4} \right) \,
\phi \left(r_{N-1} \, \sigma_{N-1} \, ,r_{N} \, \sigma_{N} \right) \,
\ee
with the quasi molecular wave function $\phi$ as a variational parameter.
The $\phi$, common to all pairs, is
$$\phi \left(r_{1} \, \sigma_{1} \, ,r_{2} \, \sigma_{2} \right) =
{1\over \sqrt{2}} \left(\uparrow\, \downarrow \, - \downarrow \, \uparrow
\right) \, \phi \left(\left|r_{1}-r_{2}\right| \right) \,. $$

It reduces to a Bose condensation of tightly bound diatomic molecules in his
limit $\xi = (k_{f} a)^{-1} \to \infty$ and to the standard BCS result in the
limit  $\xi \to - \, \infty$. This $\phi$ is well mimicked in the harmonic
oscillator representation by our $\Psi$ of eq.~(\ref{Psi}), which is made up
out of the trap harmonic oscillator wave functions and, therefore, preserves
the degeneracy in a major shell.

\vskip 0.5cm

{\bf Acknowledgments.\,\,}

\vskip 0.5cm

We thank Edward Shuryak for many
stimulating discussions, for posing semi-infinite number of provocative
suggestions and questions. We are also grateful to Hal Metcalf.
This work was partially supported by the US-DOE
grant DE-FG02-88ER40388.

\end{narrowtext}
\end{document}